\documentclass{emulateapj}
\usepackage{epsfig}
\usepackage{graphicx}

\shorttitle{Three-dimensional Doppler Images}
\shortauthors{Agafonov, Sharova, \& Richards}

\begin{document}

\title{Three-Dimensional Doppler Images of the Disk-like and Stream-like States of U Coronae Borealis}
\author{Michail Agafonov} 
\affil{Radiophysical Research Institute (NIRFI), 25, B.Pecherskaya St., Nizhny Novgorod, 603950, Russia}
\email{agfn@nirfi.sci-nnov.ru}

\author{Olga Sharova}
\affil{Radiophysical Research Institute (NIRFI), 25, B.Pecherskaya St., Nizhny Novgorod, 603950, Russia}
\email{shol@nirfi.sci-nnov.ru}

\and
\author{Mercedes Richards}
\affil{Department of Astronomy \& Astrophysics, Pennsylvania State University,
525 Davey Laboratory, University Park, PA, 16802, USA}
\email{mrichards@astro.psu.edu} 

\begin{abstract}
The 3D Radioastronomical Approach to Doppler tomography has been used to study the H$\alpha$ emission sources in U Coronae Borealis. These 3D tomograms provide greater resolution than the projected 2D version and highlight the jet-like gas flows in the $V_z$ direction transverse to the orbital plane.  In this paper, the 3D tomograms are compared at two distinct epochs when U CrB was in the disk-like state (1993 data) and the stream-like state (1994 data).  Both states display a prominent emission source, the circumprimary bulge, which is produced when the gas stream strikes the photosphere of the mass-gainer.   This source is detected within $V_z$ = $\pm$150 km s{$^{-1}$}, and demonstrates that the bulge is not confined to the orbital plane although it achieves maximum strength near  $V_z$=0 km s{$^{-1}$}.   Other emission sources include the stream-star and stream-disk shocks and a Localized Region (LR) where the circling disk material strikes the incoming gas stream.  The LR has $V_z$ velocities of 200 to 500 km s{$^{-1}$} in the disk-like state. The disk emission is seen over a range of $V_z$ velocities, and there is evidence that the disk is inclined to the orbital plane or may have two arms.  The gas stream flows along its predicted trajectory in the stream-like state, and a comparison with the disk-like state suggests that the gas stream has a higher density than the disk in both states of this binary. 
\end{abstract}


\section{Introduction}
The two-dimensional Doppler tomograms of the Algol-type binaries (\citealt{richardsetal95, albright+richards96,richards04}) display a diverse range of circumstellar structures.  These include a gas stream, accretion annulus, transient accretion disk, shock regions and sometimes a chromospheric emission source in the short-period (P $<$ 5 days) Algols, as well as a classical accretion disk in the long-period (P $\ge$  5 -- 6 days) Algols.  Some systems, like U Coronae Borealis (U CrB) and U Sagittae (U Sge), alternate between stream-like and disk-like states and display changes that can occur within a year, and sometimes overnight \citep{richards+albright99}.  These 2D Doppler tomograms are created by assuming that only one $V_z$ velocity value is represented, namely, $V_z$ = 0 km s{$^{-1}$}.    However, since the 2D map is a superposition of all the slices, then all of the Vz velocity information is embedded in this integral 2D map.  The primary aim of this work is to demonstrate that 3D Doppler tomograms can be generated by assuming a grid of $V_z$ values and that the resulting 3D maps are consistent with the 2D Doppler tomograms.  

The short-period Algols represent the class of {\it direct impact binaries} in which the gas stream from the $L_1$ point strikes the photosphere of the mass gaining star directly.  This occurs because the radius of the mass gainer is large relative to the binary separation, so the gas stream is not deflected sufficiently by the Coriolis force to avoid contact with the star \citep{richards+albright99}.  Two-dimensional hydrodynamic simulations of direct impact binaries, like U CrB, have illustrated how the high speed ($\sim500 -  600$ km s{$^{-1}$}) gas stream is shocked and heated by the impact with the slowly rotating ($v\sin i \le 150$ km s{$^{-1}$})  mass gainer \citep{blondinetal95,richards+ratliff98}.   Some of the gas from the impact/shock region is released into orbits that may repeatedly splash off the star until energy losses due to these collisions cause the gas to be accreted onto the star \citep{smak89}.   The dynamics of the impact \citep{ulrich+burger76,richards92} suggest that the shock should generate gas motions transverse to the orbital plane.   \citet{richards+ratliff98} simulated both the Cartesian and velocity evolution of the gas distribution in direct impact systems and created synthetic 2D Doppler tomograms to compare with the 2D tomograms derived from observed spectra.  They found that the change between the disk-like and stream-like states could be explained by a change in the mass transfer rate in the binary.  Three-dimensional hydrodynamic simulations of the $\beta$ Lyr eclipsing binary by \citet{bisikaloetal00b} found jets moving perpendicular to the orbital plane.  So, substantial disk structure transverse to the orbital plane can form naturally in these binaries. 

The extension to 3D Doppler tomography was first formulated in the classical paper by \citet{marsh+horne88}, and more recently discussed briefly by \citet{steeghs04} and \citet{marsh05}, who suggested that ``out-of-plane motions'' needed to be included to properly explain the observed spectra.  However, the extension to the $V_z$ direction remained unexplored until the work of \citet{agafonovetal06}, which outlined a new procedure for creating 3D Doppler tomograms.   The first 3D Doppler tomogram of a binary was constructed by \citet{agafonovetal06} from forty-seven H$\alpha$ spectra of the Algol-type binary U CrB.   They used  the reconstruction method called the Radioastronomical Approach (RA), which was developed by \citet{agafonov04a,agafonov04b,agafonov+sharova05a,agafonov+sharova05b} for the case when there is a small number of  projections (called few projections tomography; see Section 3.1). The 3D tomography of U CrB displayed  an enhanced image of the gas stream along the ballistic trajectory and an equatorial emission source centered on the primary star that were more distinct than the image produced by standard 2D Doppler tomography \citep{richardsetal95,richards01}.  A high-velocity gas jet with a high emission intensity was also discovered moving in the direction transverse to the orbital plane of the system with a high $V_z$ velocity component. This jet could not have been detected previously in the standard 2D tomograms because all velocity information is projected onto the ($V_x$,$V_y$) plane only, and hence the $V_z$ components cannot be represented in a 2D map. 

In Cartesian coordinates, a ``jet'' is an elongated feature formed by gas motions within the structure.  In this work, a jet is a luminous (high intensity) feature with the following properties: (1) It can be elongated in the ($V_x$,$V_y$) velocity planes that have $V_z$ components close to zero.   (2) The $V_x$ and/or $V_y$ velocity values of the feature should be significantly different from zero, since all ($V_x$,$V_y$) planes (for all values of $V_z$) are associated with the rotation of the binary.  (3) At the same time, the jet feature might not be elongated in the ($V_x$,$V_y$,$V_z$) velocity space if it has a high $V_z$ velocity component since this feature would  move in a direction transverse (or with significant inclination) to the direction of the orbital plane of the system.   So, it is reasonable to call the gas stream from the L1 point a jet even if it has small $V_z$ velocities.  Other jet-like features could have velocities in the $V_z$ direction. In general, any gas flows in the ($V_x$,$V_y$) plane at $V_z$=0 should move in planes that are collinear with the ($x$,$y$) orbital plane. Moreover, since gravitational fields typically dominate the gas flows, the gas velocities in the ($V_x$,$V_y$) plane at $V_z$=0 should naturally coincide with the Cartesian (x,y) orbital plane of the binary.  So, reasonable arguments can be made about the nature of the flow in the orbital plane based on the flow behavior in the ``central velocity plane'' defined by $V_z$=0, even though the velocity fields of the composite emission sources are not well-known. 

In this paper, we have continued our effort to understand the 3D structure of accretion flows by:  1) creating a procedure for constructing the 3D Doppler tomogram from the 1D projections, 2) constructing the 3D tomograms and identifying the optimum way to display these tomograms, 3) identifying the features in the 3D images and relating them to those seen in the 2D tomograms, and 4) providing an initial physical interpretation of these features.   In addition, we have used 3D tomography to study the variability of the gas flows of U CrB in its disk-like and stream-like states.  A lot of attention was focussed on the fine structure extraction of the tomogram and the analysis of the features of U CrB when it was in the disk-like state.  The visualization and analysis used in the 3D Doppler imaging process have been advanced in this work by using the RA reconstruction procedure. It is now possible with our improved understanding of the 3D velocity images to make preliminary interpretations of the 3D accretion flows.  The study of 3D Doppler tomograms of several binaries in a variety of states will help us to gain a better understanding of the significance of these transverse flows and their role in the accretion process. 

\section{2D Doppler Tomography of U CrB} 
The Algol-type binary U Coronae Borealis is an interacting system in which the cooler G0 III-IV secondary star is transferring mass to its B6 main-sequence companion (the primary) through Roche lobe overflow. U CrB is a short-period Algol (P=3.4522 days) with an orbital inclination of 79.1$^\circ$ \citep{cesteretal77} and mass ratio $q$=0.29 \citep{batten+tomkin81}. The masses of the primary and the secondary are equal to 4.8 and 1.40 $M_\odot$ and their radii are equal to 3.0 and 4.6 $R_\odot$, respectively. The systemic velocity of the system is $V_o$= -6.7 km s{$^{-1}$} and the velocity semi-amplitude of the primary is $K_p$ = 59.7 km s{$^{-1}$}. One hundred and thirty-nine H$\alpha$ (6562.8 \AA) spectra with a resolution of  0.093 \AA/pixel  or 4.3 km s{$^{-1}$} were collected in 1993 with the NSO -- McMath-Pierce telescope, and forty-seven spectra with a resolution of 0.166 \AA/pixel  or 7.6 km s{$^{-1}$} were collected in 1994 with the KPNO 0.9m Coud{\'e} Feed Telescope (see \citealt{richards+albright99}). The spectra were collected at closely spaced positions around the entire orbit of the binary.  Since the observed spectra are dominated by the spectrum of the B6 primary star, the spectral contributions of the stars were subtracted from the observed spectra to create difference spectra (see \citealt{richards04} for details). In this procedure, model atmospheres calculations were used to represent the photospheric contributions of the stars.   The difference spectra, which display enhanced emission and absorption contributions from the circumstellar gas, were processed for image reconstruction.

The earlier standard 2D Doppler tomograms of U CrB 1993 (in the disk-like state) and U CrB 1994 (in the stream-like state) taken from \citet{richards01} are shown in Figure 1. These 2D images were compared with the 3D Doppler tomograms based on the same data. The 1993 tomogram shows the presence of the following features \citep{richards01}: an accretion disk around the primary star, the gas stream, the stream-star impact region, a circumprimary emission source, and an absorption zone (region dominated by absorption). However, Figure 1 shows that the contrast of these separate features is not high because these sources are superimposed on each other in the 2D velocity image. The standard 2D Doppler tomogram reflects the summed emission on the ($V_x$,$V_y$) plane and is somewhat distorted by dispersed rays of weak emission, which are a direct consequence of the filtered back-projection method. The second frame of Figure 1 shows the gas stream along the predicted gravitational trajectory as well as circumprimary emission.  However, these features are merged instead of appearing as distinct sources.

\section{3D Doppler Imaging} 
The reconstruction of the 3D Doppler tomogram of U Coronae Borealis based on the 1993 H$\alpha$ line profiles as well as the identification of emission features are presented in this section.  This information has been used to provide new information about the binary.

\subsection{Reconstruction and Visualization of the 3D Tomogram}

The Filtered Back-Projection Method applied in computed tomography was first described in the classical paper by \citet{bracewell+riddle67}.  It had wide use in the solution of a variety of tomography problems and it was also applied to astronomical applications.  It is now widely used in Doppler tomography for image reconstruction.   Many Doppler maps have been created using the Maximum Entropy Method (MEM) in which the deconvolved image is passed through a series of statistical tests to ensure that it is consistent with the measured data (e.g., \citealt{marsh+horne88}).   The MEM then selects one image from the most feasible set of images that describe the data equally well.  However, the simpler back-projection method has been used to create Doppler images of cataclysmic variables and Algol binaries (e.g., \citealt{kaitchucketal94,richardsetal95}, and others).    The standard version of this back-projection technique maps the $V_x$ and $V_y$ components only and produces images in 2D velocity space. However, the gas flows in the directions away from the orbital plane may be significant for many binaries.  The development of the 3D version of Doppler tomography has now permitted the study of accretion structures beyond the orbital plane by incorporating the $V_z$ velocity component of the moving gas flows. 

The 3D reconstruction applied in this work is called the Radioastronomical Approach (RA).  It was developed by \citet{agafonov+sharova05a, agafonov+sharova05b}.  The theoretical possibility of image reconstruction from projections was described by J. Radon in 1917 \citep{radon1917}.  However, the mathematical equations of the Radon transform and its inverse are formulated for the ideal case with infinite numbers of projections and infinitely high resolution, covering the entire range of the frequency domain.  The widely-used Filtered Back Projection (FBP) method is based on the work of \citet{bracewell+riddle67} and permits  the application of a finite number of projections; with the image reconstruction occurring over a wide range of  spatial frequencies in the 2D Fourier domain. Equations in \citet{bracewell+riddle67} can be used to estimate the number of projections required to produce the best reconstructed image and includes the filtering of redundant high spatial frequencies.  The more general Radioastronomical Approach is based on a different set of initial conditions than FBP. It permits 3D tomographic reconstruction (see \citealt{ agafonov+sharova05b} for details).  This method is especially useful when the number of projections is relatively small because it can achieve the same resolution as FBP while using fewer profiles.   The RA can be applied in the usual cartesian coordinate space to solve a variety of  problems.   So, in the case of Doppler tomography the main distinguishing features of the RA procedure are: (1) the solution of  the {\it inverse problem} in 3D velocity space, (2) the introduction of the summarized image and summarized point spread function (SPSF), and (3) the deconvolution of the image using the  {\sc{clean}} technique to remove the responses from the sidelobes of the SPSF.  The summarized 3D image and summarized 3D point spread function were calculated based on the line profiles and one-dimensional (1D) transfer functions by incorporating the angles corresponding to the orbital inclination and orbital phases. The required version of the 3D reconstruction corresponds to the 3D$_{\rm 1D}$ variant described by \citet{agafonov+sharova05a} and  \citet{agafonov+sharova05b}. The SPSF was calculated on the set of knife-beam 1D transfer functions, which determine the velocity resolution of the profiles. This 3D$_{\rm 1D}$ notation represents the 3D reconstruction based on the 1D profiles. Here, the knife-beam function is the function or beam that transfers the intensity from the 3D figure to the 1D profile or  backwards from the 1D to the 3D image (see \citealt{ agafonov+sharova05a}). Thus, the RA reconstruction is significantly different from all other reconstruction schemes associated with computed tomography and currently applied to medicine, since those techniques derive the 3D image from the set of 2D slices and the reconstruction of these 2D slices is based on the FBP method  (e.g., \citealt{kalender05}).

In the reconstruction of the 3D Doppler tomograms, we analyzed the same set of H$\alpha$ line profiles that were used by \citet{richards01} to construct the standard 2D Doppler tomogram.  The 3D Doppler tomogram was calculated as a set of numerical values in the cells of a cube, with dimensions $(V_x,V_y,V_z)$ having values ranging from -700 to +700 km~s$^{-1}$.  The maximum of this function was normalized to the unit to permit the visualization of the image based on any slice taken from any direction.   To simplify the physical situation, the ($V_x$,$V_y$) slices are displayed in the horizontal plane (to be consistent with the direction of the  orbital plane of the binary), and the slices containing the $V_z$ axis are in the vertical direction. 

Figure 2 displays 15 slices in the ($V_x$,$V_y$) plane for values of $V_z$ ranging from $-420$ km~s$^{-1}$ to $+420$ km~s$^{-1}$, in 60 km~s$^{-1}$ steps.  The superposition of the color and contour images allows us to emphasize the main features in each $V_z$ slice of the 3D image.   Figure 3 displays enlarged images of the four most interesting $V_z$ slices. Other slices at positive $V_z$ values are shown in Figures 4 and 5, and several important cross-sections in the ($V_y$,$V_z$) plane are presented in Figures 6 and 7.  Figures 5 and 7 are included here to show the 3D slices produced from the 1994 spectra by \citet{agafonovetal06}.  Figure 8 illustrates the comparison between the distinct disk-like and stream-like distributions derived from the 1993 and 1994 spectra.   In Figures 2 -- 8, the blue contours represent absorption while red contours represent emission.  Finally, Figure 9 represents the proposed model of the U CrB binary.

\subsection{Description and Interpretation of Emission Features}
Both Figures 2 and 3 demonstrate that the most prominent emission in the UCrB (1993) image arises from the accretion disk around the primary star, as expected from the 2D image.  The disk emission strengthens as we move from $V_z$= $-420$ km~s$^{-1}$ towards the $V_z$= 0 km~s$^{-1}$ slice and then weakens again by $V_z$= +420 km~s$^{-1}$.  This $V_z$ velocity range will decrease with the adjustment for the lower resolution in the $V_z$ direction.  The disk emission has an asymmetrical structure, as though composed of two parts with different velocities. Figure 2 shows that the maximum intensity of the mass gaining star occurs near the central $V_z$ velocities (near $V_z$=0). However, the maximum emission intensity in the 3D tomogram ($I_{max}$ = 1.0) was found in the $V_z$= $-20$ km~s$^{-1}$ slice (Figure 3), and corresponds to the velocity location of the primary star.  The intensities of the mass loser and mass gainer are aligned in a symmetrical way along the $V_x$=0 line. The intensities of these sources was strongest near $V_z$=0, decreased in both directions towards $V_z$= $\pm$180 km~s$^{-1}$, and was negligible at higher absolute $V_z$ velocities.

We consider ten bright emission features that are distinctly visible in the tomograms.   Most of these features were identified earlier in the 2D Doppler tomograms, and from comparisons between the predicted locations of these structures in both velocity-space and Cartesian space. The ten features are outlined in Table 1 along with their typical velocity ranges.  Eight of these features are identified in Figure 3, which shows the four most interesting slices corresponding to $V_z$ = $-240, -20, 240$ and $420$ km~s$^{-1}$ in the 1993 tomogram.  The features are (1) circumprimary emission and accretion annulus, (2) chromospheric emission associated with the of cool secondary donor star, (3) the stream-disk impact region (hot-spot) in the middle of the predicted path of the stream, (4) the stream-star impact region, where the gas stream strikes the stellar surface, (5) the locus of  a Keplerian  accretion disk, (6) a bright feature near the center of the tomogram,  previously identified as the Localized region by \citet{richards92}, (7) another feature which may be associated with the Localized Region, and (8) a feature near the velocity of the mass loser, but with high $V_z$ velocities.  The sources labeled 3, 6, 7, 8 dominate in brightness at the higher positive $V_z$ velocities. These sources are illustrated in Figure 4, for slices from $V_z$ =240 to 540 km~s$^{-1}$.   Two additional sources were identified in the 1994 tomogram of \citet{agafonovetal06} and are shown in Figure 5: (9) the gas stream flowing along the predicted ballistic trajectory, and (10) a high velocity stream (or jet) moving in the transverse direction relative to the orbital plane of the binary.  

The velocity resolutions in the ($V_z$,$V_x$) plane and the ($V_z$,$V_y$) plane depend on the orbital inclination of the binary and are equal only for an inclination of 45$^\circ$. For high values of inclination, the resolution in the $V_z$ direction is not as good as in the other dimensions, and gets worse as the values of $V_z$ increase.  Since the orbital inclination of U CrB is high ($i$ = 79.1$^\circ$), the 3D tomogram was restored with a resolution of 20 km~s$^{-1}$ in the $V_x$ and $V_y$ directions and 70 km~s$^{-1}$ in $V_z$ direction. So, the features on the tomogram look elongated in the $V_z$ direction because the resolution in the $V_z$ direction is approximately 3.5 times lower than in the $V_x$ and $V_y$ directions (see Figures 6 and 7).   The full dimensions of all features are given in Table 1 and throughout the text. 

The possibility of an adjustment for the lower resolution in the $V_z$ direction depends on whether a source is compact or elongated.  A source is considered to be compact if it has one expressed intensity maximum and a narrow range of velocities, while an extended source is characterized by a wide range of velocities in more than one direction ($V_x$,$V_y$ or $V_z$). For example, the circumprimary emission (Feature {\bf 1}) is obviously compact because it has only one intensity maximum and a narrow range of velocities.  Its isolines (equal intensity lines) have circular shapes because of equal velocities in the $V_x$ and $V_y$ directions (Figure 2 or 3).  Feature {\bf 1} is artificially elongated in the $V_z$ direction (in Figure 7, right frame) only because of the poorer resolution in that dimension.  So, the measured $V_z$ velocities for compact sources should be scaled by the factor of 1/3.5 (= 0.29) before direct comparison with the $V_x$ or $V_y$ velocities. The deconvolution procedure is more complicated for extended sources (e.g., Features 6 - 8), so any features with the ratio $V_z$/$V_x$ or  $V_z$/$V_y$ greater than $3.5$ are considered to elongated in the $V_z$ direction. Consequently, the shapes of the isolines in the ($V_y$,$V_z$) or ($V_x$,$V_z$) planes can be compared with those in the ($V_x$,$V_y$) plane once we take into account the factor of 0.29 scaling of the artificially stretched shapes of all the components in the $V_z$ direction.  

The compact blue absorption feature seen in the lower left quadrant of Figure 2 from $V_z$ = -180 to +420 km~s$^{-1}$  (see also Figure 8, lower right frame) was called the {\it absorption zone} by \citet{richards01,richards04},  and is associated with the locus where the gas temperature becomes too high to emit at H$\alpha$.  A structure at a similar location was identified in the ultraviolet spectrum of U Sge by \citet{kempner+richards99}. 

A short description of each feature is provided below.    Figures 2 -- 7 have been used to examine and interpret the features.   

\noindent
1.)  The {\it circumprimary emission} (centered on the mass gainer) and an {\it accretion annulus} (a circular region of low velocity) are labeled ({\bf 1}) in Figure 3.   These structures are centered along the $V_y$ axis ($V_x$ = 0 km~s$^{-1}$) and slightly below the $V_x$ axis in all the slices from $V_z$= $-120$ to $+120$ km~s$^{-1}$. The full extent of this structure is approximately 360 km~s$^{-1}$ in the $V_z$ direction, and ranges from about $V_z$= $-180$ to $+180$ km~s$^{-1}$ (see Figure 2), while its width is only 120 km~s$^{-1}$  in the ($V_x$,$V_y$) plane.  Since the resolution in the $V_z$ direction is 3.5 times worse than that in the ($V_x$,$V_y$) plane (Section 3.1), the corrected $V_z$ width should be (360 km~s$^{-1}$/3.5) $\simeq$100 km~s$^{-1}$ (or $V_z$=-50 to $V_z$= +50 km~s$^{-1}$), which is comparable to the width of the structure in the ($V_x$,$V_y$) plane.  This result suggests that the circumprimary structure is nearly spherically symmetric in the velocity domain.

\noindent
2.) {\it Chromospheric emission} from the cool donor star (the secondary) is shown at ({\bf 2}). Its $V_z$ velocity ranges from $V_z$= $-120$ to $+120$ km~s$^{-1}$, as shown in Figure 2.

\noindent
3.) A {\it bright (hot) spot} in the middle of the gas stream is located at ({\bf 3}) and represents the site where the gas stream makes impact with the outer edge of the accretion disk. It is detected in the slices from $V_z$= $+180$ to $+540$ km~s$^{-1}$.  However, emission along the predicted gravitational trajectory of the gas stream looks faint in all slices based on 1993 spectra.  This stream emission is clearly brighter in the 1994 tomogram (see Figure 5).  Feature {\bf 3} may also represent gas ejected out of the orbital plane by the impact with the disk.  This ejected gas has high positive $V_z$ velocities from 240 to 540 km~s$^{-1}$, centered on 420 km~s$^{-1}$, near $V_x$ = -360 km~s$^{-1}$  (see Figure 6 and the slice for Vz=420 km~s$^{-1}$ in Figure 3).  It is apparent that the ejected gas then flowed within the predicted locus of the accretion disk, along the trajectory marked in Figure 3 (bold dashed line), and with a lower $V_z$ velocity of 240 km~s$^{-1}$.   As this gas blends with the disk (Feature {\bf 5}), it acquires negative $V_z$ velocities up to -200 km~s$^{-1}$.    This transition is evident in Figure 3 (slice $V_z$= -240 km~s$^{-1}$).

\noindent
4.) The location of the {\it stream-star impact region}, where the gas stream strikes the stellar surface, is found at ({\bf 4}) near the location where the gas stream trajectory intersects the inner part of the disk. Its velocity ranges from $V_z$= $-300$ to $+60$ km~s$^{-1}$.  

\noindent
5.) Beyond the impact/splash zone, the gas motion at ({\bf 5}) is consistent with the {\it predicted locus of the accretion disk} and rotates like a disk.  Its $V_z$ velocities ranged from $-420$ to $+420$ km~s$^{-1}$ (see Figures 2 -- 4), so this disk has substantial flow in the vertical direction, although the velocity range will be slightly smaller after the adjustment for the lower resolution in the $V_z$ direction.   This result was expected from studies of spectra collected during primary eclipse.  This disk may be turbulent with vortical movements, especially if it is inclined.   We can identify the stream-disk and stream star regions even when the disk emission is prominent if the disk has a lower density than these regions or if it is inclined to the orbital plane of the binary.  Some evidence in favor of an inclined disk is provided by Figure 2, in which the lower left part of the disk is visible mainly for negative $V_z$ slices and the emission seems to shift to the lower right part of the disk for positive $V_z$ slices.  In addition, one part of the stream (flowing along the gravitational trajectory) makes impact with the disk, and subsequently, the bright feature {\bf 3} (hot spot along the stream) has positive velocities (so it is moving away from us). At the same time, the stream-star impact region (Feature {\bf 4}) has negative $V_z$ velocities, so this transverse region is moving towards us (Figure 6).  The emission intensity of the disk is not as strong as some of the other features, but it is evident that the gas flow in the disk spreads around the primary star.  Moreover, we cannot exclude the existence of two arms within the disk with different sets of velocities, as was predicted  by \citet{sawadaetal86} and \citet{spruit87} and then observed in the case of the cataclysmic variable IP Peg by \citet{steeghsetal97} and others.

\noindent
6 \& 7.) The two concentrated regions labeled ({\bf 6}) and ({\bf 7}) are found near the center of the tomogram in the same general region that was identified as the {\it Localized Region} in Cartesian models of $\beta$ Per \citep{richards92,richards93}, RW Tau \citep{vesper+honeycutt93}, and other direct-impact systems. These regions are detected at high positive $V_z$ velocities ranging from $+240$ to $+540$ km~s$^{-1}$.  At the highest $V_z$ velocity, they become part of the same structure. As explained by \citet{richards92}, this overall structure is created by the natural flow within the disk when it circles the primary and strikes the incoming gas stream, to produce a disk-stream shock region. A similar structure consisting of jets moving perpendicularly to the orbital plane was also identified from hydrodynamic models of the eclipsing binary, $\beta$ Lyr A \citep{bisikaloetal00b}.  These regions can be examined in greater in detail in Figure 4, which shows fragments of six slices corresponding to high positive $V_z$ up to $+540$ km~s$^{-1}$.  They can also be seen in Figure 6 which shows the ($V_y$,$V_z$) cross-sections.  This figure shows that features {\bf 6} and {\bf 7} have approximately the same $V_z$ velocities ($V_z$ $\sim$400 km~s$^{-1}$) as feature {\bf 3} (which represents the stream-disk impact in the middle of the stream).  Feature {\bf 3} is seen in slices $V_x$= -360 and -300 km~s$^{-1}$ while Feature {\bf 6} is shown in slices $V_x$= -20 and 0 km~s$^{-1}$.  Although Features {\bf 3} and {\bf 6} have similar $V_z$ velocities, they have different $V_x$ velocities;  the gas velocity vector for Feature {\bf 6} is almost perpendicular to the central velocity plane (and hence the orbital plane), while Feature {\bf 3} has two velocity components ($V_x$ and $V_y$) similar to the disk, its  $V_z$ velocity component has a high value of $\sim$400 km~s$^{-1}$.  As a result, it is reasonable to conclude that Feature 3 is the part of the disk gas that was ejected out of the orbital plane after the impact with the stream.  The regions {\bf 6} and {\bf 7} could explain why the disk gas breaks up into clumps and changes its velocity direction near the site of the Localized region where the circling disk gas makes impact with the trajectory of the gas stream \citep{richards92,kuznetsovetal01,bisikaloetal00a}. 

\noindent
8.) The feature labeled ({\bf 8}) has a velocity similar to the $V_x$ and $V_y$ velocities of the donor star, but with high positive $V_z$ velocities in the range from $V_z$=$+200$ to $+540$ km~s$^{-1}$ (see Figures 2 -- 6).  This structure may be chromospheric in origin but the high velocities suggest that it could correspond to a jet or flare associated with the chromosphere of the secondary star.  However, it is also likely that features {\bf 6}, {\bf 7}, and {\bf 8} may all be related to the interaction between the material in the disk and the inner edge of the stream.  The $V_z$ = 480 km~s$^{-1}$ slice in Figure 4 hints at a connection between regions {\bf 7} and {\bf 8}, although they seem to be separate in other slices.  The low velocity gas that circles the star may merge with the disk when it interacts with the gas stream (to become Features 6 \& 7, and perhaps {\bf 8}), or it could escape from the system in the $V_z$ direction. 

\noindent
9.) The feature labeled ({\bf 9}) represents the gas stream flowing along its predicted ballistic trajectory.  It flows from the L1 point towards the mass gaining star with speeds up to 450 km~s$^{-1}$.  This feature is seen in Figure 5 for the U CrB (1994) data \citep{agafonovetal06}.

\noindent
10.) The feature labeled ({\bf 10}) is the high-velocity stream or jet.  It is a compact source in the ($V_x$,$V_y$) plane but has a significant transverse flow in the ($V_y$,$V_z$) plane, with negative  $V_z$ velocities primarily from $-400$ to  $0$ km~s$^{-1}$ \citep{agafonovetal06}. It can be seen in Figures 5 and 7 for the U CrB (1994) data.  Gas speeds of up to 600 km~s$^{-1}$ were expected at the star stream impact site from dynamical arguments presented by \citet{ulrich+burger76,richards92}.

\subsection{Variability of the Gas Flows in U CrB}

The 3D Doppler tomogram of the U CrB binary based on the 1993 data (hereafter, 3D-1993) displays an extensive accretion disk, a circumprimary emission source, a chromospheric emission source and a jet-like structure associated with the secondary star, as well as stream-star and stream-disk shock regions.  \citet{albright+richards96} suggested that the binary was in a {\it disk-like state} at this epoch.   It is useful also to compare these results with those derived from the 1994 data (hereafter, 3D-1994).   

The 2D tomogram of U CrB based on 1994 data (Figure 1) shows that the gas flow in the binary was in a {\it stream-like state} during that epoch \citep{richardsetal95}. This 2D image is dominated by the gas stream which flows along the predicted gravitational trajectory from the L1 point towards the mass-gaining star. \citet{agafonovetal06} used 3D Doppler tomography to produce a more distinct image of the gas stream than possible with the standard 2D method.  The cascade visualization of the 3D-1994 image is shown in Figure 5, which demonstrates that most of the gas stream flow (Feature {\bf 9}) was in the central planes close to $V_z$ = 0.  It is obvious that the motion of any such gas must will be in the central velocity plane (and also the orbital plane) if it has no $V_z$ velocity component.  A transverse flow of gas (a gas jet, Feature {\bf 10}) was also found with a high $V_z$ velocity component and two other weak components ($V_x$=-60 and $V_y$=0 km~s$^{-1}$; Figure 7, left).  It is probable that a feature with such velocity components may have arisen from the expected location of the stream-star impact site \citep{agafonovetal06}.

The cascade visualization of the 3D-1994 image is shown in Figure 5, which demonstrates that most of the gas stream flow (Feature {\bf 9}) was in the central velocity plane defined by $V_z$ = 0.   It is obvious that the motion of any such gas must will be in the central velocity plane if it has no $V_z$ velocity component.  A transverse flow of gas (a gas jet, Feature {\bf 10}) was also found with a high $V_z$ velocity component and two other weak components ($V_x$= $-60$ km~s$^{-1}$ and $V_y$=0 km~s$^{-1}$; Figure 7, left).  It is probable that a feature with such velocity components may have arisen from the expected location of the stream-star impact site \citep{agafonovetal06}.   The ($V_y$,$V_z$) cross section of this gas jet is shown in Figure 7 (left) for $V_x$= $-60$ km~s$^{-1}$, adapted from \citet{agafonovetal06}. Figure 7 (right) also shows the ($V_y$,$V_z$) cross section (at $V_x$= 0 km~s$^{-1}$) of the emission source which corresponds to an equatorial bulge on the primary star taken from the same 3D-1994 image. 

The ($V_y$,$V_z$) cross sections for the 3D-1993 image (Figure 6) can be compared with the 3D-1994 results shown in Figure 7.  The center of the equatorial emission on the primary star (feature {\bf 1}) is displaced at $V_z$ = 80 km~s$^{-1}$ towards positive $V_z$ velocities in 3D-1994, but the same feature is located near the center in 3D-1993. The displacement is only of -20 km~s$^{-1}$, which is not significant. However, the differences in the velocity structure of the transverse flows in the two states of the binary may be significant.   

The ($V_y$,$V_z$) cross section for $V_x$ = $-60$ km~s$^{-1}$ for 3D-1994 shows the transverse flow. The velocity of the transverse flow (Feature {\bf 10}) in 1994 (Figure 7) ranges from $V_z$ = $-400$ to $0$ km~s$^{-1}$, and the flow is nearly perpendicular to the orbital plane.   The gas stream impact site (Feature {\bf 4}) in 3D-1993 has a similar velocity range at the end of the stream in the slice for $V_x$= -580 km~s$^{-1}$, but this gas does not flow in a perpendicular direction across the orbital plane. The correspondence between Features {\bf 4} and {\bf 10} is reasonable since this jet may have resulted from the splash at the impact site (see \citealt{agafonovetal06}).   However, the other transverse gas flows for 3D-1993 have the opposite direction relative to similar flows in 3D-1994. Feature {\bf 3} in the middle of the stream and features {\bf 6}, {\bf 7}, and {\bf 8} range from $V_z$ = 200 to 500 km~s$^{-1}$ (see Figure 6).  This is the main difference in the behavior of the transverse flows at the two epochs. 

The high velocity transverse jet (Feature 10; Figure 5) and the localized region (Feature 6; Figure 4) have similar locations in the ($V_x$,$V_y$) plane, but the cross sectional views show that these features move in different directions (see Figures 6 and 7) in the ($V_y$,$V_z$) plane.     Feature 6 has positive velocities with $V_z$ $\sim$ 400 km~s$^{-1}$, and is flowing away from us, while Feature 10 has negative velocities with $V_z$ $\sim$ $- 200$ km~s$^{-1}$, and is flowing towards us.   However, it is interesting that Feature 10 (the high velocity transverse jet; Fig. 5b) and Feature 4 (the star-stream impact region; Fig. 5a)  have the same velocity range from $V_z$ =  $-400$ to $0$ km~s$^{-1}$, which confirms the conclusion by \citet{agafonovetal06} that the high velocity jet is located at the site of the impact between the gas stream and the star.

It should also be noted that the range of $V_z$ velocities for feature {\bf 9} (gas stream) is nearly symmetric about the $V_z$=0 direction in the 3D-1994 images (Figure 5).  We also notice that the gas stream or jet (feature {\bf 10}) flowing in $V_z$ direction has negative $V_z$ velocities,  but later the velocities become positive suggesting motion of the gas  around the binary.  As a result, the circumprimary emission in 1994 has the displacement of the maximum intensity to $V_z$=80 km~s$^{-1}$ (in positive values of $V_z$). This fact was noted in the stream-state of the primary. 

While the image reconstruction process can now produce 3D velocity images, we are still unable to create the desired 2D or 3D Cartesian images because the various emission features have different velocity fields.   However, an estimate of the size of the accretion disk can be made from information provided in Figure 1. Based on the point where the gas stream is truncated, we can produce a Cartesian model to show the dominant emission sources in U CrB.  The distance from the L$_1$ point to the center of the primary star has been divided into 10 equal segments in the center frame of Figure 1, and the similar small circles on the Doppler tomograms are marked at velocity intervals corresponding to those distance intervals.   Comparing the center and right frames of Figure 1, we see that the gas stream flow is truncated after traveling 6.5 small circles on the gas stream trajectory, which is equivalent to 65\% of the distance from the L$_1$ point to the center of the primary star.  The stream-star interaction occurs along the path of the gas stream right where the gas stream ends (at 65\% of the distance from the L$_1$ point).  However, Figure 1 (left) shows that the flow within the disk does not touch the star, since the gas within the disk (Feature {\bf 5}) flows in the middle of the predicted locus of the disk.  In addition, since the gas flow overlaps with the dashed circle in the tomogram, which corresponds to the Keplerian velocity at the Roche surface of the primary star, then the disk must extend out to the physical Roche surface of the star.  The radius of the star, $R_p$ = 3.0 R$_\odot$, $q$ =0.29, the binary separation, $a$ = $1.23 \times 10^7$ km, and the Roche lobe radius, $R_{L1} = a (0.5 + 0.227 \times log10 (q^{-1})$ = 11.0 $R_{\odot}$ = 3.7 R$_p$.  So, the inner edge of the accretion disk has a radius of $\sim$1.3$R_p$, while the outer  edge of the disk has a radius of $\sim$3.7$R_p$.  Both disk radii are measured from the center of the primary.

A model of the U CrB binary is shown in Figure 9.  This figure explains the various features detected in the 3D tomograms of this binary.    In the stream-like state (Fig. 9a), the geometry displays the case of direct impact without a disk and shows a luminous gas stream from the donor to the mass gainer.  The 3D tomography results suggest that the stream impact site is displaced to the lower part of the star.  In that case (see Fig. 9b), it is reasonable that the splash and reflection of the gas stream (Feature {\bf 10} or F10) can achieve moderate velocity components ($V_x \sim -60$ km~s$^{-1}$, $V_z \sim -200$ km~s$^{-1}$; red lines).  Beyond this splash site, the gas flows back towards the star (blue lines) and as a result,  Feature {\bf 1} (F1 - annulus) has a positive displacement in $V_z \sim 80$ km~s$^{-1}$  (see also Fig. 7 ).   In the disk-like state (Fig. 9c), the geometry displays the case of direct impact with a disk but no intense luminous gas stream from the donor to the mass gainer  (see also Figs. 2 and 6).  The gas stream then interacts with the external part of the disk  and the halo (gas located between the stars) and this interaction with the weak stream creates Feature {\bf 3} (F3) with velocity components ($V_x \sim -360$ km~s$^{-1}$, $V_z \sim 400$ km~s$^{-1}$).  However, the disk (and halo) decelerates or brakes and ejects gas with velocity components ($V_x \sim 80$ km~s$^{-1}$, $V_z \sim 400$ km~s$^{-1}$.  The ejected gas can leave the system or fall back toward the star (and so it has negative $V_z$ velocities).   This is similar to the dynamical simulation results of \citet{kuznetsovetal01} (their Fig. 3a and Fig. 4).  The remaining lower part of the gas stream strikes the disk but is too dilute to have a significant impact on the disk.  This interaction is shown in Feature {\bf 4} (F4), with velocity components ($V_x \sim -580$ km~s$^{-1}$, $V_z \sim -200$ km~s$^{-1}$).  The $V_z$ component for F4 is identical to that for F10, but the $V_x$ component is $Ð580$ km~s$^{-1}$ because this represents the location where the stream ends, or perhaps because the stream impact occurs with the inner part of the disk, which has this value of $V_x$.  The result is the formation of an annulus without any displacement in $V_z$.  

\section{Conclusions}

The dynamical effects of gas flows perpendicular to the orbital plane and the vertical structure of the gas stream were considered in the classic theoretical work by \cite{lubow+shu76}.  Our 3D Doppler tomography results reveal that the process of mass transfer is a more complex phenomenon than expected. The accumulation of additional results from 3D Doppler tomography of different binaries will be useful in producing a better understanding of the multiple emission sources in these binaries and how the transverse flows influence the accretion process.   The visualization of the features with the aid of  3D Doppler tomography represents a new step in data processing.  It is especially useful in the identification and study of the transverse flows as well as features in the orbital plane.  The comparison of  the disk-like and stream-like states is illustrated in Figure 8 by means of a volume representation of the main central slices ($V_z$ = 0) for 3D-1993 and 3D-1994 and also by means of color and contour images in this plane of the binary.   It is clear that the disk-like state of U CrB in 1993 is  distinct from the stream-like state in 1994. 

The main results of this work are (1) the application of a new tool to produce 3D Doppler tomograms, (2) the visualization of the disk-like and stream-like states of the Algol-type U CrB interacting binary, and (3) the discovery of gas flows moving across the orbital plane and the identification of the strongest emission sources (Features 1-10) in the 3D Doppler tomogram.

\noindent
1. {\it The application of a new tool to produce 3D Doppler tomograms.} The 3D Doppler tomograms created by using the Radiastronomical Approach contain more information than the images produced with the standard 2D Filtered Back Projection Method.    This new RA method represents a qualitative advance in data processing.  The cascade of 2D images along the $V_z$ direction have now provided a new understanding of the gas flows in the direction transverse to the orbital plane.  This is true even without correcting for the artificial elongation of the sources in the $V_z$ direction because of the lower resolution in that direction.

\noindent
2. {\it The visualization of the disk-like and stream-like states of the binary.}  The 2D Doppler tomograms of U CrB constructed from H$\alpha$ line profiles collected in 1993 and 1994 show two different states of the binary:  a transient accretion disk representing the disk-like state (in epoch 1993) and a bright gas stream flowing along predicted ballistic trajectory representing the stream-like state (in epoch 1994). These emission sources are even more distinct in the 3D Doppler tomograms, and both states show evidence of a circumprimary emission source.   Emission from the chromosphere of the cool secondary star is also visible especially in 1993 when the binary was in the disk-like state.  The multiple slices of the 3D tomogram show that the emission from the transient accretion disk is visible in the $V_z$ direction from $V_z$= -420 to +420 km~s$^{-1}$, although this range will be slightly smaller after the adjustment for the lower resolution in the $V_z$ direction. The detection of negative $V_z$ velocities as well as positive $V_z$ velocities associated with the disk emission suggests that the disk may be inclined to the direction of the orbital plane (the $x$,$y$) plane  or we may be observing two arms of the disk, as observed in the cataclysmic variable IP Peg by \citet{steeghsetal97}. 

The disk emission is not as strong as some of the other features.  Moreover, the disk emission in the disk-like state is significantly weaker than the gas stream emission in the stream-like state.   These observations suggest that the disk is not as dense as the gas stream or the other compact features, and may explain why the tomogram of U CrB in its disk-like state does not show a symmetric disk which fills up its predicted locus in the ($V_x$,$V_y$) plane.  The fact that the gas stream emission can dominate at certain epochs is further evidence that the gas stream is denser than the disk.  These conclusions can be confirmed by comparing the hydrodynamic calculations of \citet{richards+ratliff98} for two Algol-type binaries, $\beta$ Per and TT Hya, with those of \citet{sawadaetal86,spruit87,kuznetsovetal01} for the cataclysmic variable, IP Peg.  Both TT Hya and IP Peg display dominant accretion disks over long timescales, while $\beta$ Per shows no accretion disk.  It is apparent that U CrB is different from IP Peg because the gas stream in U CrB is denser than the disk.  This is why we can see the shock regions within the disk.  However, in IP Peg, it seems that there is no gas stream, so the disk dominates the image so you can see the completed circular pattern for the disk within the predicted locus of the disk.  That is definitely not true for U CrB because the top part of the disk is missing, probably because the denser gas stream prevents the disk from completing the trip around the star.

\noindent
3. {\it The discovery of transverse gas flows and the identification of the several emission sources.}  The reconstruction of the 3D image of U CrB has clearly demonstrated that there exist strong gas flows which have high $V_z$ velocities.  Therefore, we can conclude that these gas flows move in directions that are distinct from that represented by the standard 2D Doppler tomogram, which implicitly assumes that $V_z$ = 0 km~s$^{-1}$.   In both states of this binary, a prominent source of emission is the circumprimary equatorial bulge on the primary star, which is produced when the gas stream strikes the surface of the star.  In the disk-like state, the 3D image displays several additional compact emission sources.  One of these is located along the gas stream trajectory, and is associated with the stream-disk interaction ($V_x$ $\sim$ -300 km~s$^{-1}$, $V_y$ $\sim$ 100 km~s$^{-1}$, $V_z$ $\sim$ 300 to 400 km~s$^{-1}$) and the other corresponds to impact/splash region where the gas stream strikes the surface of the primary star ($V_x$ $\sim$ -580 km~s$^{-1}$, $V_z$ $\sim$ -300 to +100 km~s$^{-1}$).  Such high velocity flows (jets) have been anticipated since the gas stream makes a high speed impact with the photosphere at speeds of up $\sim$500-600 km~s$^{-1}$, compared to the slower rotation of the mass gainer \citep{richards92, blondinetal95, richards+ratliff98}.  A second pair of bright features is found near the center of the ($V_x$,$V_y$) plane at $V_z$ velocities of $\sim$ +200 to +500 km~s$^{-1}$. These are found in a location similar to the Localized Region in $\beta$ Per, RW Tau, and other direct-impact systems.   

In the stream-like state, there is evidence of a strong transverse flow with $V_z$ velocity components of up to -300 km~s$^{-1}$ (see Fig. 7). This gas jet may arise as a result of the stream-star interaction and overlaps with the location of this region in the 3D tomogram of U CrB when it was in the disk-like state.  We have also found that the transverse flows associated with the circumprimary emission in the disk-like state flow in the opposite direction than when the binary is in the stream-like state.  In the disk-like state, the flows have luminous $V_z$ velocity components in both positive and negative directions.  Figure 6 shows that  Features 3, 6, 7, and 8 have positive $V_z$ components while Feature 4 has negative $V_z$ velocities.   However, in the stream-like state, the transverse flow has a high negative $V_z$ velocity of -300 km~s$^{-1}$ (see Feature {\bf 10} in Figure 7 for the $V_x$= -60 km~s$^{-1}$  slice), while the portion with positive $V_z$ velocities reaches a value of only $\sim$100 km~s$^{-1}$ (see Feature {\bf 1} in Figure 7 for the $V_x$ = 0 km~s$^{-1}$ slice).   This suggests that the dense gas stream may penetrate through the rarified disk material and strike the star to form the circumbinary structure as an extension of the photosphere, but some of this material spreads out of the orbital plane, preferentially in the direction away from the observer, and  at nearly seven times the synchronous velocity of the star.  However, in the disk-like state, the disk material is denser than before and so the gas stream does not have as strong an impact on the stellar surface.  As a result, the gas is spread more evenly in both directions out of the orbital plane.  Another explanation is that when the disk material circles the star and comes into contact with the denser gas stream, the interaction produces Features {\bf 6} and {\bf 7} and displaces the stream-disk site at {\bf 3}, on the outer part of the disk, so that the jet flows in the opposite direction.  This explanation is consistent with the existence of an inclined accretion disk, and so we cannot rule out the possibility that the rotation axis of the primary star may not be perpendicular to the orbital plane, as expected, and could be precessing.  

The calculation and analysis of 3D Doppler tomograms of additional systems will enable us to make a significant qualitative jump in the study of accretion in interacting binaries. The reconstruction procedure based on the Radioastronomical Approach will be influential in this future work. 

\acknowledgements
This research was partially supported by Russian Foundation for Basic Research (RFBR) grant 06-02-16234, and a grant from The Pennsylvania State University.

\begin{figure*}
\figurenum{1}
\center
\includegraphics[height=60mm]{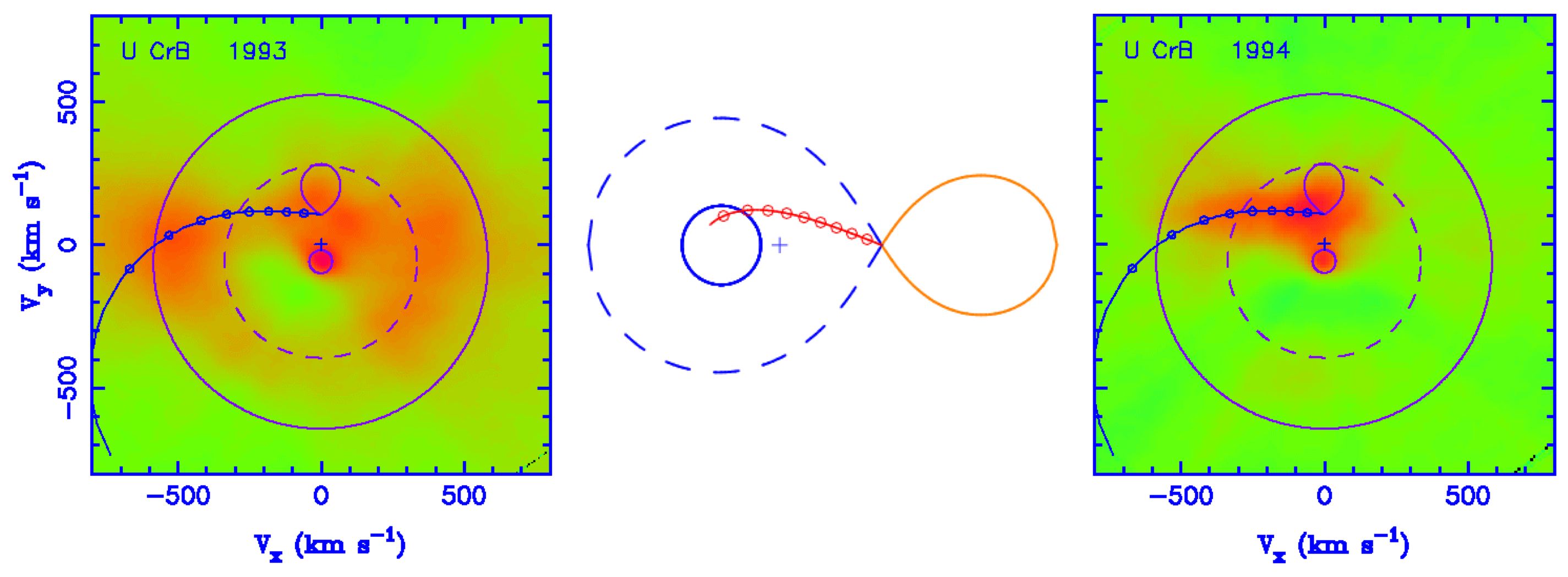}
\caption{2D Doppler tomograms of U CrB-1993 in the disk-like state (left) and U CrB-1994 in the stream-like state (right) taken from \citet{richards01}, and a Cartesian model of the binary (center).
}
\label{f1}
\end{figure*} 

\begin{figure*}
\figurenum{2}
\center
\includegraphics[height=200mm]{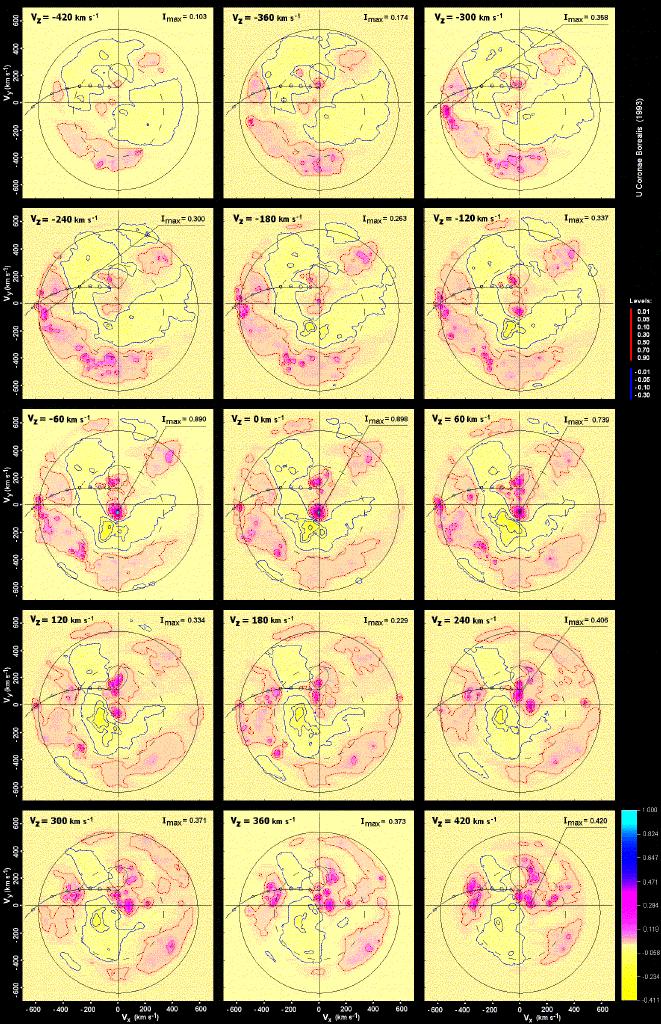}
\caption{Slices of the 3D Doppler tomogram of  U CrB from data collected in 1993. The images are shown for intensity levels: + 0.01, 0.05, 0.1, 0.3, 0.5, 0.7, 0.9 (red) and -0.01, -0.05, -0.1, -0.3 (blue), with velocities beyond the orbital plane of $V_z$ = -420 km~s$^{-1}$ to $V_z$ = +420 km~s$^{-1}$ in steps of 60 km~s$^{-1}$.
}
\label{f2}
\end{figure*} 

\begin{figure*}
\figurenum{3}
\center
\includegraphics[width=165mm]{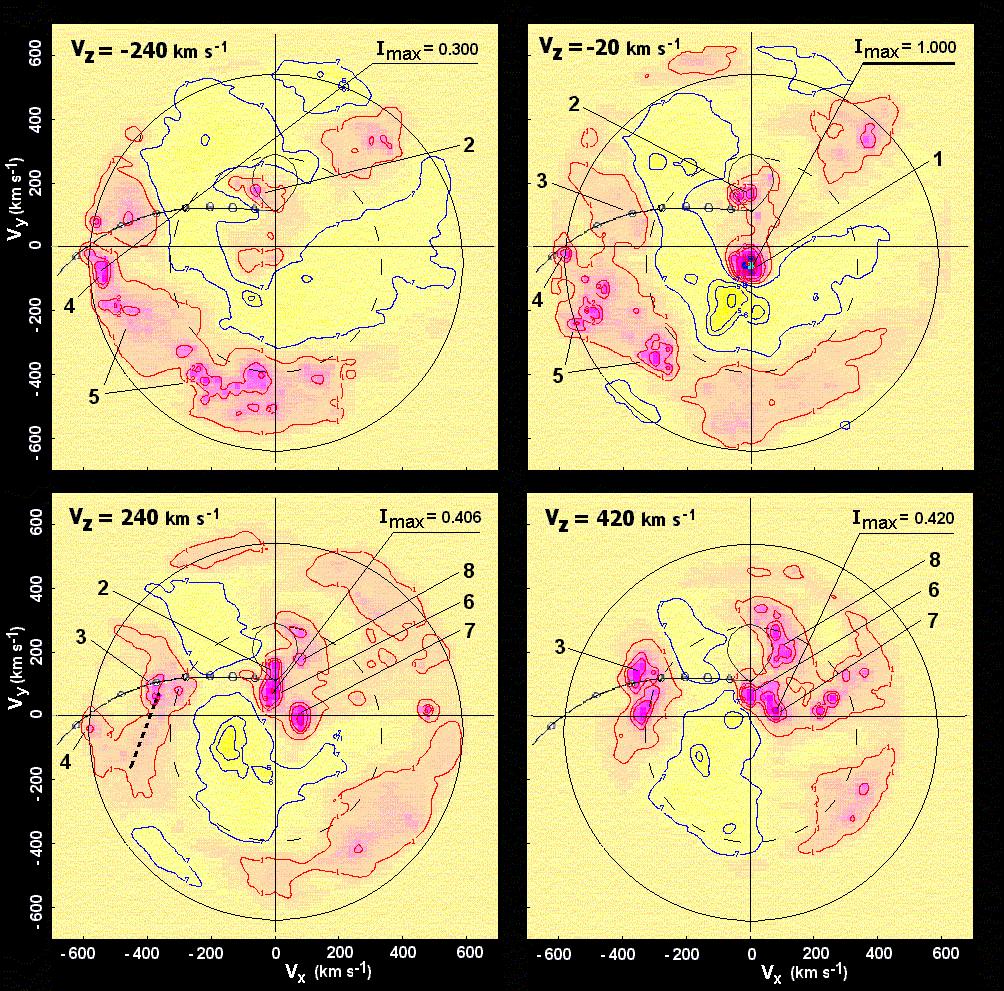}
\caption{Four slices in the ($V_x$,$V_y$) plane for $V_z$ = -240, -20 (maximum intensity), 240, 420 km~s$^{-1}$ showing the locations of the interesting features labeled 1-8. The levels of emission intensity in contour and color are the same as earlier in Figure 2.
}
\label{f3}
\end{figure*}

\begin{figure*}
\figurenum{4}
\center
\includegraphics[height=180mm]{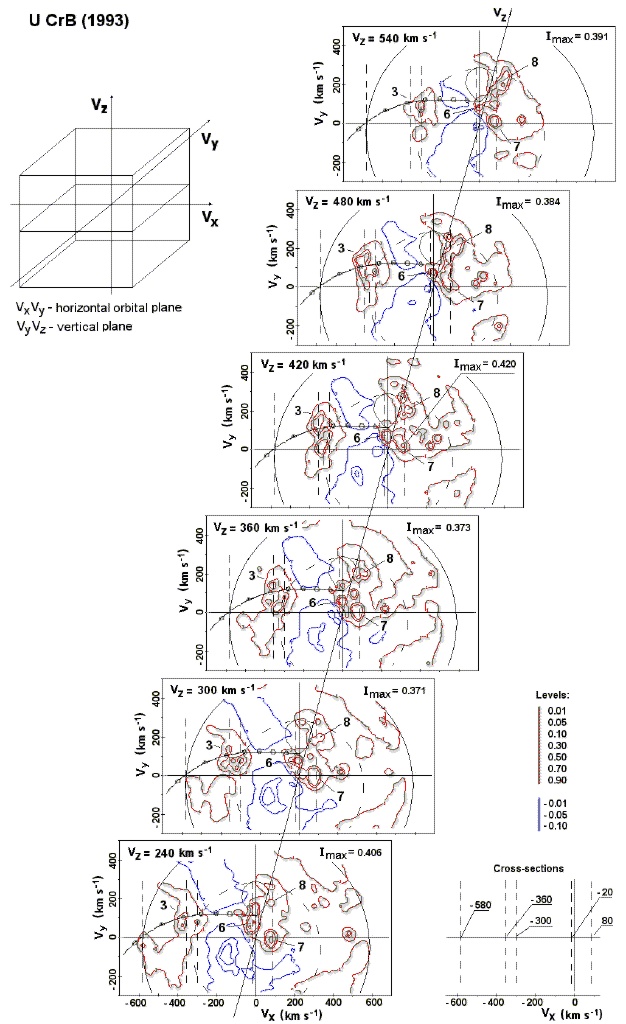}
\caption{ The 3D Doppler tomogram of U CrB-1993 showing a cascade visualization of the ($V_x$,$V_y$) slices from $V_z$ = 240 to 540 km~s$^{-1}$ which display the interesting features labeled 3, 6, 7, 8.
}
\label{f4}
\end{figure*} 

\begin{figure*}
\figurenum{5}
\center
\includegraphics[height=180mm]{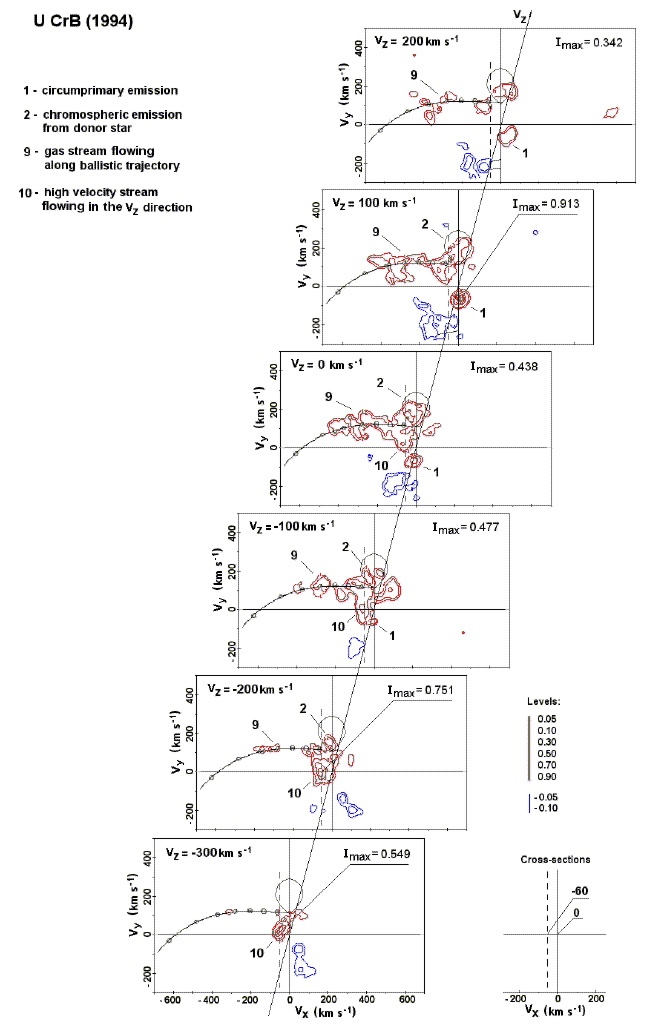}
\caption{The 3D Doppler tomogram of U CrB-1994 showing a cascade visualization of the most interesting slices (adapted from \citet{agafonovetal06}).  The resolution in the $(V_x,V_y,V_z)$ directions correspond to velocities $(30 \times 30 \times 110)$ km~s$^{-1}$.  The three strongest H$\alpha$ emission features are (1) the accretion annulus, (2) chromospheric emission, (9) the gas stream flowing along the ballistic trajectory from the L1 point, and (10) the high velocity stream flowing in the $V_z$ direction. 
}
\label{f5}
\end{figure*} 

\begin{figure*}
\figurenum{6}
\center
\includegraphics[width=165mm]{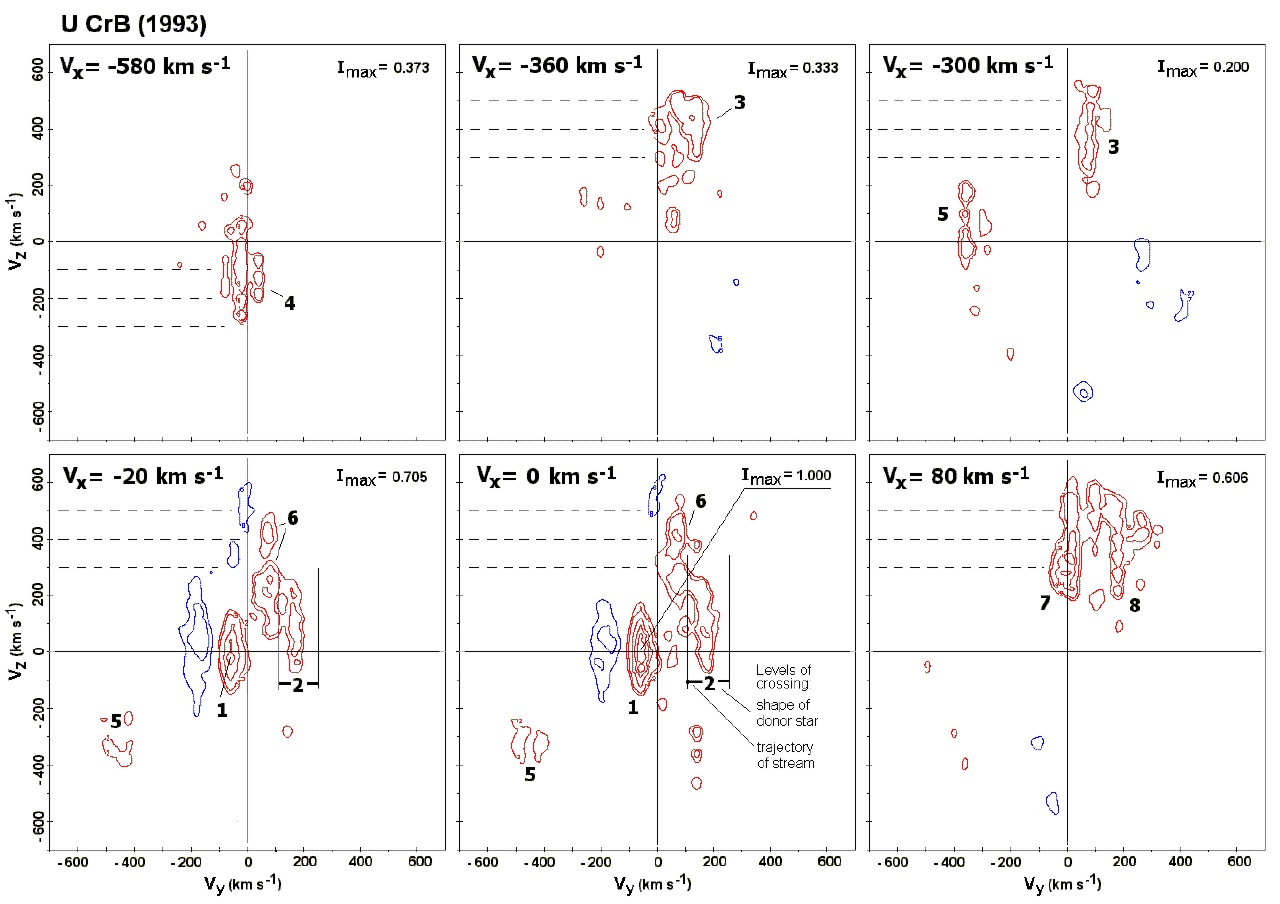}
\caption{Cross sections in the ($V_y$,$V_z$) plane in the 3D tomogram of U CrB -1993 for $V_x$ = -580, -360, -300, - 20, 0, 80 km~s$^{-1}$.   The red contours correspond to the emission (positive intensities 0.05, 0.1, 0.3, 0.5, 0.7, 0.9) while the blue contours correspond to the absorption (negative intensities from -0.05 to -0.1). 
}
\label{f6}
\end{figure*} 

\begin{figure*}
\figurenum{7}
\center
\includegraphics[width=165mm]{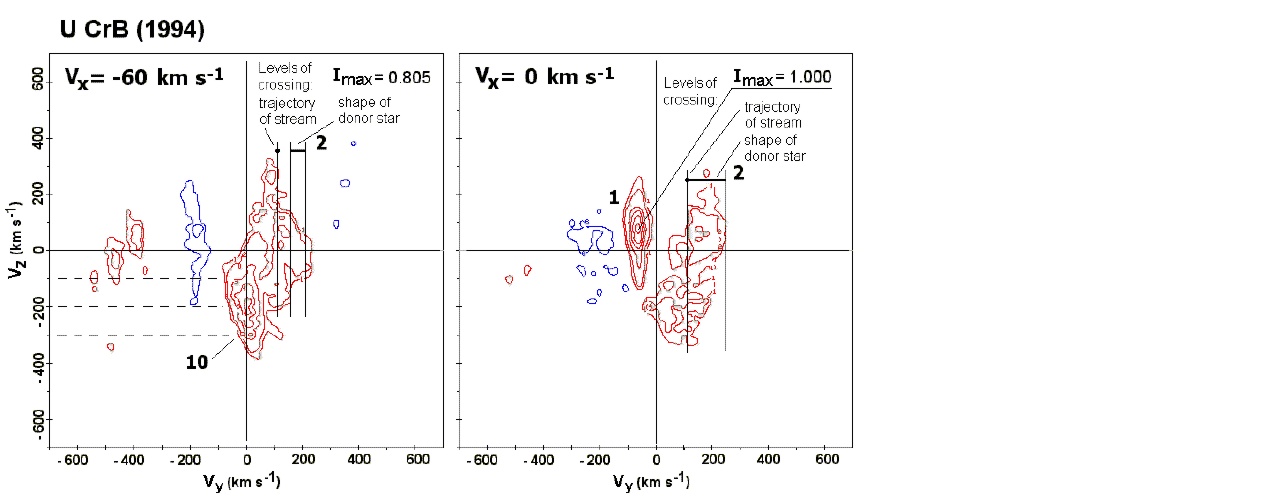}
\caption{Cross sections in the ($V_y$,$V_z$) plane in the 3D tomogram of  U CrB-1994 for $V_x$ = -60 and  0 km~s$^{-1}$ adapted from \citet{agafonovetal06}. The right frame illustrates emission from the equatorial bulge on the primary star. The left frame shows the transverse flow within the jet  (10) as material is ejected from orbital plane away from the location where the gas stream strikes the surface of the mass gainer. The red contours correspond to the emission (positive intensities 0.05, 0.1, 0.3, 0.5, 0.7, 0.9) while the blue contours correspond to the absorption (negative intensities from -0.05 to -0.1). 
}
\label{f7}
\end{figure*}

\begin{figure*}
\figurenum{8}
\center
\includegraphics[width=165mm]{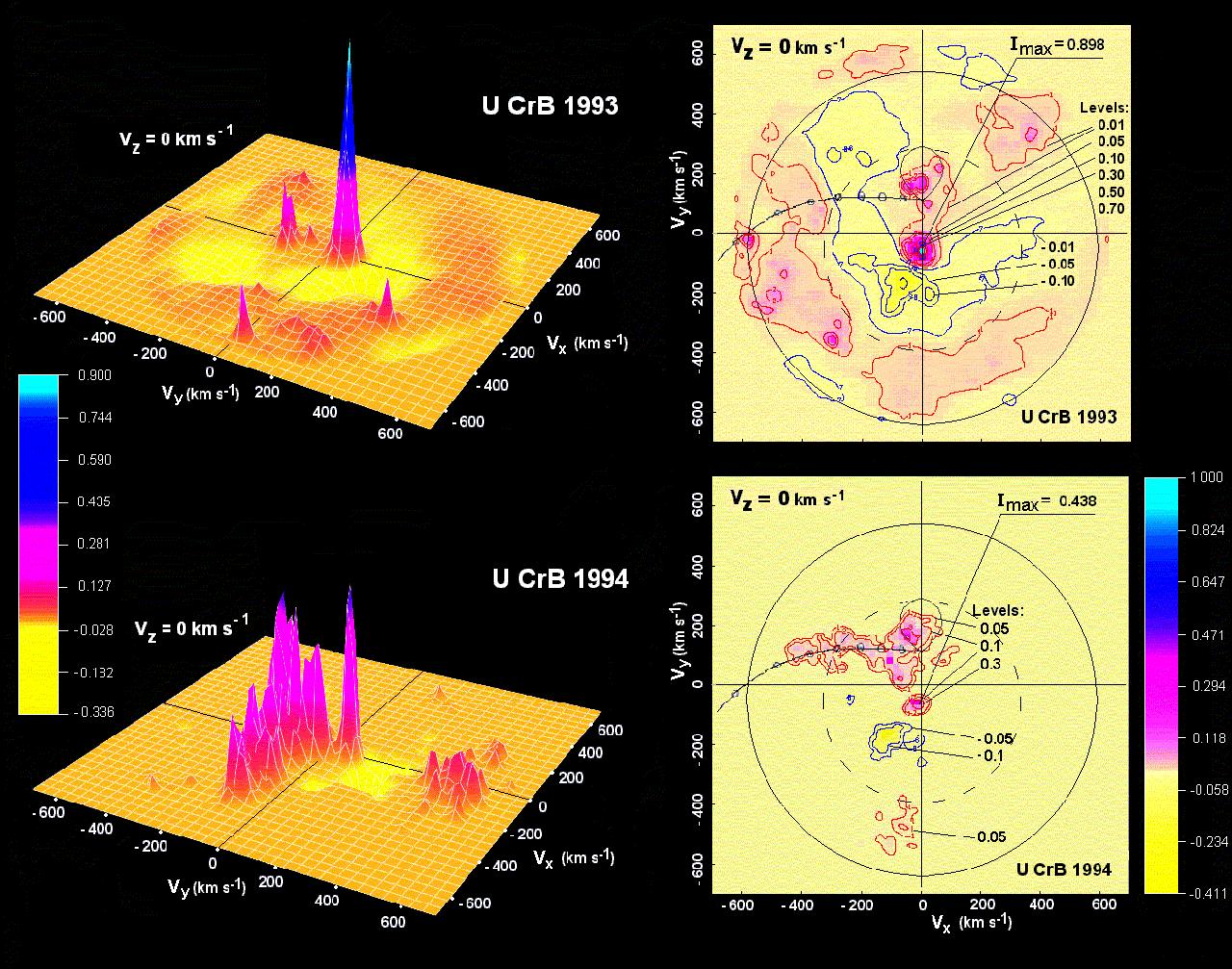}
\caption{Volume representation of the central slices ($V_z$=0) and also their color+contour representation on the plane of the 3D Doppler tomogram calculated from the U CrB-1993 data compared to that found from the U CrB-1994 data  (from \citealt{agafonovetal06}). See intensity levels on the color scales. 
}
\label{f8}
\end{figure*}
\clearpage

\begin{figure*}
\figurenum{9}
\center
\includegraphics[width=165mm]{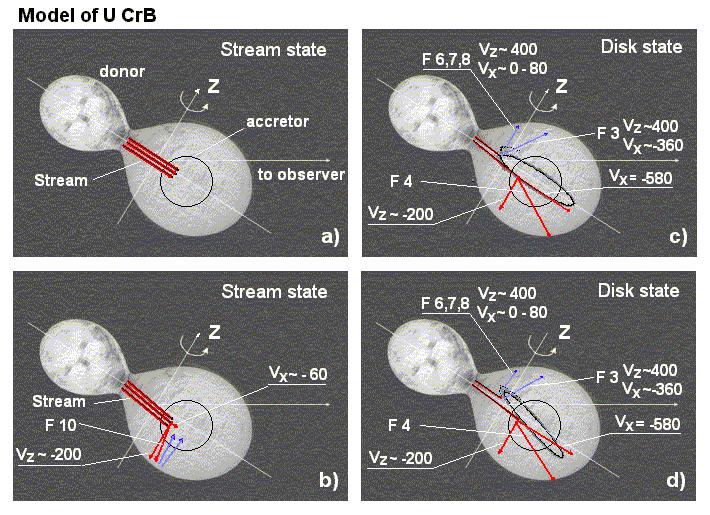}
\caption{A proposed model of the U CrB binary in Cartesian ($x$,$z$) coordinate space: a) the stream-like state with the central direct stream impact on the mass gainer and without the disk; b) the displaced stream impact on the lower part of the mass gainer in the form of a jet (F10; red line) and also  gas falling onto the star (blue line); c) the disk-like state with a weak stream, which interacts with the disk. Features 3, 6, 7, 8 are the result of the diskÐstream interaction after the disk gas circles the star, but Feature 3 has a higher $V_x$ velocity than the other features.  The lower part of the stream interacts only with the inner part of the disk or with the photosphere of the star  to produce feature F4, which is a jet with two velocity components ($V_x \sim -580$ km~s$^{-1}$ and $V_z \sim200$ km~s$^{-1}$  below the disk).  Finally, d) the same as in (c) but with the disk inclined.  
}
\label{f9}
\end{figure*}

\clearpage

\begin{deluxetable}{clccc} 
\tablecolumns{5} 
\tablewidth{0pc} 
\tabletypesize{\scriptsize}
\tablecaption{Characteristics and locations of prominent emission features in the 3D tomogram} 
\tablehead{  
\colhead{} & \colhead{ } & \multicolumn{3}{c}{Location: central velocity or velocity range (km~s$^{-1}$)} 
} 
\startdata
{No.} & {Emission Feature} & {$V_x$} & {$V_y$}  & {$V_z$} \\
\hline
1 & Circumprimary emission & 0 (-60 to + 60)  & -60 (0 to -120) & -20  (-180 to 180) \\
2 & Chromospheric emission & -80 to 80 & 120 to 230 & -360 to 240 \\
3 & Stream-disk impact region & -360 (-300 to -400) & 100 (0 to 200) & 420 (60 to 540) \\
4 & Stream-star impact region & -580 (-520 to -600) & 0 (-100 to 100) & (-300 to +60) \\
5 & Locus of the Accretion disk &  {\tablenotemark{a}} &  {\tablenotemark{a}} & -420 to +420 \\
6 & Localized Regions -- Part 1 & -20 (-60 to +30) & 75 (30 to 120) & 200, 400 (150 to 540) \\
7 & Localized Regions -- Part 2 & 75 (40 to 110) & 0 (-50 to +50) & 300 (200 to 540) \\
8 & Feature near donor star & 80 (70 to 160) & 200 (130 to 270) & 360 (200 to 540) \\
9 & Gas stream flow & {\tablenotemark{b}} & {\tablenotemark{b}} &   \\
10 & High velocity transverse jet & -60  & 0 &  -400 to 0 \\
\enddata
~~~\tablenotetext{a}{corresponds to the predicted location of the feature on the 2D ($V_x$,$V_y$) plane.}
~~~\tablenotetext{b}{corresponds to the predicted ballistic trajectory of the gas stream.}
\end{deluxetable}

\end{document}